\documentclass[aip,pop,reprint,amsmath,amssymb]{revtex4-2}
\usepackage{graphicx}
\usepackage{bm}
\usepackage{silence}%
\usepackage[colorlinks=true,allcolors=blue]{hyperref}
\makeatletter\let\write@bibliographystyle\relax\let\auto@bib@innerbib\relax\let\auto@bib\relax\makeatother

\newcommand{\B}{\mathbf{B}}
\newcommand{\x}{\mathbf{x}}
\newcommand{\Phit}{\Phi_t}
\newcommand{\Phip}{\Phi_p}

\begin{document}

\title{Exact force-free three-dimensional toroidal equilibria and the sawtooth \texorpdfstring{$q(0)$}{q(0)} problem}

\author{Wrick Sengupta}
\email{wsengupta@princeton.edu}
\affiliation{Department of Astrophysical Sciences, Princeton University, Princeton, NJ 08543, USA}
\author{Amitava Bhattacharjee}
\affiliation{Department of Astrophysical Sciences, Princeton University, Princeton, NJ 08543, USA}
\author{Taylor J. Klotz}
\affiliation{Department of Applied Mathematics, University of Colorado, Boulder, CO 80309, USA}
\author{Andrew O. Brown}
\affiliation{Department of Astrophysical Sciences, Princeton University, Princeton, NJ 08543, USA}
\affiliation{Princeton Plasma Physics Laboratory, Princeton, NJ 08540, USA}

\date{\today}

\begin{abstract}
We construct exact, closed-form solutions of $\nabla\times\B=\lambda\B$ with nested toroidal flux surfaces around a closed magnetic axis, finite rotational transform and, in general, no continuous Euclidean symmetry. They are counterexamples to Grad's conjecture in its pressureless form. Inside a regular flux surface that fits in a ball of radius $r_b$ with $|\lambda|r_b<4.49$, each is the unique minimum of magnetic energy at fixed toroidal flux and helicity, so each is a Taylor state. The rotational transform is a ratio of one-dimensional integrals. Our primary application is the sawtooth $q(0)$ problem in tokamaks. In one configuration, $q$ crosses unity on a smooth torus of closed field lines, with no island and no current sheet. Across the family, minimum-energy fixed-boundary states with $q(0)<1<q_{\rm wall}$ exist; in our scans $q(0)$ reaches $0.84$ at tight aspect ratio and about $0.94$ at conventional aspect ratio. At comparable helicity-to-flux ratio, configurations with different boundary shapes have either sign of the magnetic shear. These fixed-boundary equilibria are candidates for relaxed cores, which need not be axisymmetric. Their dynamical accessibility and their matching to the surrounding plasma remain open.
\end{abstract}

\maketitle

\paragraph*{Introduction.}
Exact solutions of the magnetohydrodynamic (MHD) equilibrium equations in three dimensions are rare, and those with nested toroidal flux surfaces are rarer still. Grad conjectured that, apart from configurations with a continuous symmetry, smooth toroidal equilibria with nested surfaces generically fail to exist~\cite{Grad1967,Grad1985}. Three-dimensional (3D) counterexamples with pressure have since been constructed~\cite{GLT2026,Landreman2026,Issan2026a,Issan2026b,DeLacySato2026,EPPT2026}. In its pressureless form ($\nabla p=0$) the conjecture asserts that a smooth force-free field with nested tori around a regular closed axis and finite rotational transform must have a continuous Euclidean symmetry. Exact nonsymmetric vacuum fields with nested tori were given in a previous paper of ours~\cite{Sengupta2026vac} and by L\'opez~\cite{Lopez2026}.

In this Letter we give exact, closed-form force-free fields with constant, nonzero $\lambda=\mu_0J_\parallel/B$, nested tori and no continuous symmetry. To our knowledge they are the first explicit examples with fully nested tori and no continuous Euclidean symmetry. Previous KAM results give Cantor families of invariant tori~\cite{EncisoPeralta2015,EncisoLuquePeralta2020,EncisoPeraltaRomaniega2023}, and admissible nonconstant $\lambda$ profiles are rare~\cite{EncisoPeralta2016,ClellandKlotz2020,Abe2022,PeraltaVaquero2024}. Taylor proposed that turbulent relaxation leads to a minimum-energy state at fixed magnetic helicity and flux, with spatially constant $\lambda$~\cite{Taylor1974,Taylor1986}. We prove that each of our fields is the unique energy minimum in any regular wall that satisfies an explicit enclosing-ball condition, and we compute its safety factor $q=1/\iota$ from one-dimensional integrals.

Our primary application is to the sawtooth $q(0)$ problem. Kadomtsev's model of the crash~\cite{Kadomtsev1975} predicts complete reconnection inside the $q=1$ surface, leaving $q(0)\geq1$. Faraday rotation on TEXTOR~\cite{Soltwisch1988} and motional Stark effect (MSE) polarimetry on TFTR~\cite{Yamada1994} reported $q(0)\approx0.7$--$0.8$ throughout the cycle. On DIII-D, ensemble-averaged MSE reconstructions gave $q(0)=0.96\pm0.08$ before and $1.02\pm0.1$ after the crash (standard deviations)~\cite{Wroblewski1993}. In an ensemble-averaged set of DIII-D sawteeth, Faraday-polarimetry-constrained reconstructions gave $q(0)$ rising from about $0.88$ to $0.93$ at the crash and remaining below unity throughout the cycle~\cite{Benedett2024}. On KSTAR, MSE gives $q(0)\approx1.0\pm0.03$ with a possible systematic offset of about $0.05$~\cite{Nam2018}, and an analysis of excited $m=2,3$ modes favors $q(0)\geq1$ after the crash~\cite{Park2016,Nam2018}. A recent nonlinear model keeps $q(0)$ slightly above unity and attributes the crash to $m=n>1$ interchange modes~\cite{Jardin2020}. Measuring $q$ near the axis remains difficult~\cite{Jardin2020,Nam2018}, and the interpretations differ.

Taylor relaxation of the core is an obvious candidate for the post-crash state. In the cylinder, the Bessel-function model has $q$ monotonically decreasing in $\psi$, so $q(0)<1$ implies $q<1$ everywhere, as in reversed-field pinches. In cylindrical geometry, Bhattacharjee \textit{et al.} obtained tokamak-like states with monotonically increasing $q$ profiles, but only by imposing additional global invariants~\cite{Bhattacharjee1980}. In axisymmetric toroidal geometry, Taylor~\cite{Taylor1975} and Bhattacharjee and Wiley~\cite{Bhattacharjee1983} found tokamak-like Taylor states with $q$ rising outward, with very weak shear ($1.13$ to $1.16$ in one example), which give way to pinch-like states as $\lambda$ grows, at $q\approx0.75$--$0.8$; no state with $q$ crossing unity inside the plasma was reported. Gimblett and Hastie matched a Taylor-relaxed core to an ideal exterior and obtained post-crash $q(0)<1$~\cite{Gimblett1994}, and the crash has been modeled as a sequence of multi-region relaxed equilibria~\cite{Qu2025}.

Here we give exact, generally three-dimensional equilibria with a smooth $q=1$ surface inside the relaxed region. They include states with $q(0)<1<q_{\rm wall}$ that need not be axisymmetric. Their occurrence depends on the shape of the wall.

\begin{figure*}[t]
\includegraphics[width=0.8\textwidth]{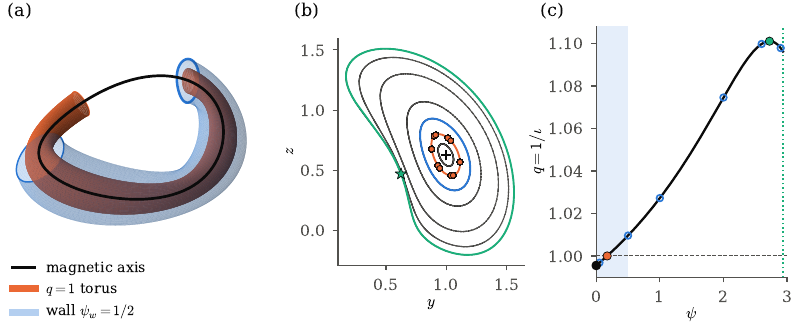}
\caption{Configuration E1 ($m_g=1/2$, $(R_0,S_0)=(0.9,0.5)$, $a=b=1$, $\lambda=9/4$). (a) The non-planar magnetic axis, the $q=1$ torus $\psi\approx0.172$ and the wall $\psi_w=1/2$. (b) The section $x=0$: exact level curves of $\psi$ and punctures of field lines traced for 600 turns; the twelve dots on the $q=1$ torus are six closed lines. (c) $q(\psi)$ from the period formula (solid curve) and from tracing (circles, differences below $2\times10^{-5}$); $q$ crosses unity on the $q=1$ torus, and the shear reverses at the maximum of $q$, $\psi\approx2.725$, beyond the wall; shaded, the tube $\psi\le1/2$; dotted, the edge of the nested region, $\psi_c\approx2.936$.}
\label{fig:E1}
\end{figure*}

\paragraph*{Exact force-free fields.}
Following Whittaker~\cite{Whittaker1903}, Bergman~\cite{Bergman1929} and Gilbert~\cite{Gilbert1960}, we construct solutions of $\nabla\times\B=\lambda\B$, $\nabla\cdot\B=0$ in the form
\begin{equation}
\B(\x)=\mathrm{Re}\,\frac{1}{2\pi i}\oint_\Gamma e^{i\lambda\rho}\,\mathbf{k}(w)\,f(\eta,w)\,dw,
\label{eq:WB}
\end{equation}
with $\mathbf{k}=(1-w^2,\,i(1+w^2),\,2w)$, $\eta=\mathbf{k}\cdot\x$ and $\rho=z-w(x-iy)$. For real nonzero $\lambda$ the integral solves the equations for any function $f$, called the associate, that is holomorphic in $\eta$ near a fixed contour $\Gamma$ avoiding its singularities. Because $\mathbf{k}$ is a null vector, i.e. $\mathbf{k}\cdot\mathbf{k}=0$, each equation $\mathbf{k}(w)\cdot\x=g(w)$ defines a complex plane, and we make $f$ singular on a one-parameter family of such planes. We choose $g(w)$ so that every plane of the family is tangent to the surface
\begin{equation}
\frac{x^2}{\ell-1}+\frac{y^2}{\ell-m_g}+\frac{z^2}{\ell}=1,\qquad 0<m_g<1 .
\label{eq:quadrics}
\end{equation}
For $\ell>1$ these are ellipsoids, for $m_g<\ell<1$ hyperboloids of one sheet, and for $0<\ell<m_g$ hyperboloids of two sheets. One surface of each kind passes through every point, so the three roots of Eq.~(\ref{eq:quadrics}) are Jacobi's orthogonal ellipsoidal coordinates. We call this family of confocal surfaces the web, and the parameter $m_g$ sets its shape. In these coordinates the integral~(\ref{eq:WB}) can be evaluated in closed form~\cite{Sengupta2026vac,Sengupta2026long}. We parametrize the three roots by Jacobi elliptic functions, $\ell=1/\mathrm{sn}^2(R|m_g)$, $\nu=m_g\,\mathrm{sn}^2(S|m_g)$ and $\mu=\mathrm{dn}^2(\tau|1-m_g)$, so that $R$, $S$ and $\tau$ label the ellipsoid, the two-sheet and the one-sheet hyperboloid through each point, and use $u=2(R+S)$, $v=2(R-S)$ and $\zeta=2\tau-K_c$, where $K_c=K(1-m_g)$. Each surface of constant $\zeta$ is one quadrant of a hyperboloid of one sheet, and $\zeta$ advances by $8K_c$ in one turn around the $z$ axis, so it plays the role of a toroidal angle. In these coordinates the field and a flux function are
\begin{equation}
\B=A_1\sin\chi_1\nabla u-A_2\sin\chi_2\nabla v-\mathcal{K}\nabla\zeta,\qquad \psi=M^2-\mathcal{K}^2,
\label{eq:field}
\end{equation}
where $A_1=a\,e^{-\lambda H}$, $A_2=b\,e^{\lambda H}$, $\chi_j=\lambda(S_j-S_{j,0})$ with $S_{j,0}$ the value of $S_j$ at $(R_0,S_0)$, $M=A_1+A_2$ and $\mathcal{K}=A_1\cos\chi_1+A_2\cos\chi_2$. The phases $S_1=F(R)-G(S)$, $S_2=F(R)+G(S)$ and the weight $H$ are closed-form combinations of Jacobi elliptic and zeta functions~\cite{Sengupta2026long}. A configuration of the family is fixed by $(m_g,R_0,S_0,a,b,\lambda)$, with $a,b>0$, where the base point $(R_0,S_0)$ is a fixed pair of values of the coordinates $R$ and $S$. We write $\Delta F=F(K)-F(R_0)$ and $\Delta G_\pm=\pm[G(\pm K)-G(S_0)]$, with $K=K(m_g)$, for the phase increases from the axis to the focal ellipse $R=K$ and to the focal hyperbola $S=\pm K$, and $\Delta_{\min}=\min(\Delta F,\Delta G_\pm)$.

\begin{figure*}[t]
\includegraphics[width=\textwidth]{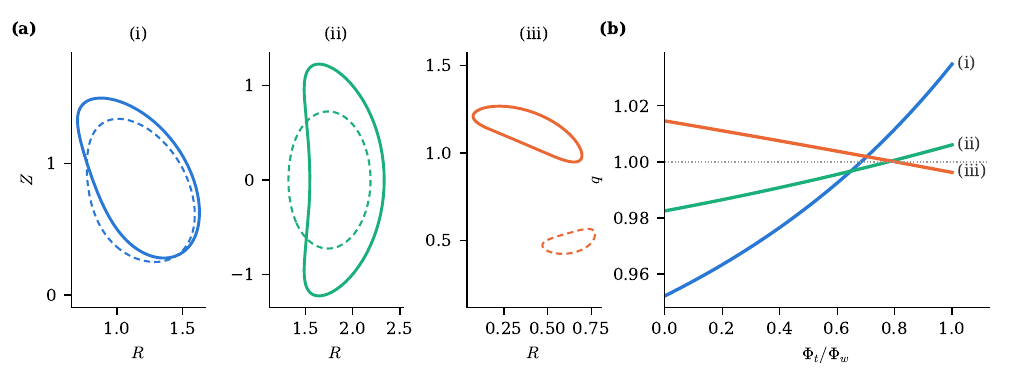}
\caption{Three configurations of the family at nearly equal helicity-to-flux ratio, $k_T=K/\Phi_w^2=1.000\pm0.003$, Eq.~(\ref{eq:kT}). (a) Wall cross-sections in the planes $x=0$ (solid) and $y=0$ (dashed). (i) $m_g=0.8$, $(R_0,S_0)=(0.3K,0.25K)$, $\lambda=1.979$, $\psi_w=1.3349$; (ii) $m_g=0.35$, $(0.3K,0)$, $\lambda=1.523$, $\psi_w=1.208$; (iii) $m_g=0.2$, $(0.6K,0.75K)$, $\lambda=3.209$, $\psi_w=0.510$; all with $a=b=1$. (b) $q$ against normalized toroidal flux. Cases (i) and (ii) have $q(0)<1<q_{\rm wall}$; case (iii) has $q$ falling outward through unity.}
\label{fig:shape}
\end{figure*}

One verifies directly that Eq.~(\ref{eq:field}) satisfies $\nabla\times\B=\lambda\B$, $\nabla\cdot\B=0$ and $\B\cdot\nabla\psi=0$. We state four properties of these fields for any amplitudes $a,b>0$; the proofs will be given in Ref.~\onlinecite{Sengupta2026long}.
\begin{list}{}{\setlength{\leftmargin}{1.6em}\setlength{\labelwidth}{1.4em}\setlength{\labelsep}{0.2em}\setlength{\topsep}{3pt}\setlength{\itemsep}{2pt}\setlength{\parsep}{0pt}}
\item[(i)] \emph{Magnetic axis.} The base-point curve $R=R_0$, $S=S_0$ is a closed magnetic axis, on which $\psi=0$.
\item[(ii)] \emph{Nested surfaces.} Every flux surface $\psi=\psi_w$ around the axis with $0<\psi_w<\psi_*=4ab\sin^2[\min(|\lambda|\Delta_{\min},\pi/2)]$ is a smooth torus enclosing the axis, so the nested region extends at least to $\psi_*$. If $\Delta F<\min(\Delta G_+,\Delta G_-)$ and $0<|\lambda|\Delta F<\pi/2$, the surfaces first touch a focal conic exactly at $\psi_*$, on the focal ellipse; if instead one of $\Delta G_\pm$ is strictly the smallest and $0<|\lambda|\Delta_{\min}<\pi/2$, they first touch the nearer focal hyperbola, again at $\psi=\psi_*$.
\item[(iii)] \emph{No continuous symmetry.} For $S_0\neq0$ the axis is not planar; for $S_0=0$ it is the ellipse $(A\,\mathrm{sn}\,\tau,\,B\,\mathrm{cn}\,\tau,\,0)$, with $\tau$ of parameter $1-m_g$ and $B^2-A^2=1-m_g>0$, which is not a circle. In either case, when nearby lines do not rotate by a multiple of $2\pi$ per turn, no continuous Euclidean symmetry preserves the axis. The axis transform is numerically noninteger for the planar-axis examples below (approximately $1.19$, $1.02$ and $1.01$ for the low-$q(0)$ configuration, Fig.~\ref{fig:shape}(ii) and the near-axisymmetric fit).
\item[(iv)] \emph{Minimum energy.} Let the wall be any regular flux surface $\psi=\psi_w$ that lies inside a ball of radius $r_b$, and let $|\lambda|<x_1/r_b$, where $x_1\approx4.4934$ is the first positive root of $\tan x=x$. Then the field is the unique minimizer of $\int B^2\,dV$ among divergence-free fields tangent to the wall with the same toroidal flux and the same helicity $K$ of Eq.~(\ref{eq:kT}). The proof~\cite{Sengupta2026long} rests on the helicity bound for fields in a ball~\cite{Cantarella2000}. Since $|\x|^2=\ell+\nu+\mu-1-m_g\le\ell=1/\mathrm{sn}^2(R|m_g)$, and every regular tube of (ii) has $R>R_{\rm low}$ with $F(R_{\rm low})=F(R_0)-\Delta_{\min}$, the ball of radius $r_b=1/\mathrm{sn}(R_{\rm low}|m_g)$ about the origin contains the wall, which gives the explicit window $|\lambda|<x_1\,\mathrm{sn}(R_{\rm low}|m_g)$ for every configuration. We checked numerically that all 7500 scanned configurations below satisfy this condition, each with $|\lambda|$ at least 0.1\% below the largest allowed value; for configuration E1 this bound is verified by interval arithmetic. In the same window the field is the only constant-$\lambda$ field with its wall, flux and $\lambda$. Taylor's bifurcation to a mixed helical state~\cite{Taylor1974} requires $\lambda$ at or above the first eigenvalue, so it cannot occur in this window.
\end{list}
These fields are therefore counterexamples to Grad's conjecture in its pressureless form.

The flux periods of Eq.~(\ref{eq:field}) separate, so the safety factor $q=1/\iota$ on each torus of (ii) is a ratio of one-dimensional periodic integrals~\cite{Sengupta2026long}. We therefore compute $q(\psi)$ without tracing field lines:
\begin{equation}
\begin{gathered}
q=\operatorname{sgn}(\lambda)\,\frac{P_R+P_S}{P_\tau},\qquad P_\tau=\int_0^{4K_c}\frac{d\tau}{\sqrt{M^2-\psi}},\\
P_R=\int_0^{2\pi}\frac{d\sigma}{2|\lambda|\sqrt{ab}\,\sqrt{1-s^2\sin^2\sigma}\,(\ell-\bar\mu)},
\end{gathered}
\label{eq:qsep}
\end{equation}
where $P_S$ is the same integral with $\bar\mu-\nu$ in place of $\ell-\bar\mu$, $s^2=\psi/4ab$, $\bar\mu=E(1-m_g)/K_c$ with $E$ the complete elliptic integral of the second kind, and $\ell$ and $\nu$ are taken at $F(R)-F(R_0)=\lambda^{-1}\arcsin(s\sin\sigma)$ and $G(S)-G(S_0)=\lambda^{-1}\arcsin(s\sin\sigma)$, respectively.

\paragraph*{A Taylor state with $q$ crossing unity.}
Configuration E1 (Fig.~\ref{fig:E1}) has $m_g=1/2$, $(R_0,S_0)=(0.9,0.5)$, $a=b=1$ and $\lambda=9/4$. Its nested region extends to $\psi_c=2.9362$. With the wall at $\psi_w=1/2$, outward-rounded interval evaluations of the flux periods give
\begin{equation}
\begin{aligned}
0.995&<q_{\rm axis}<0.996,\\
1.008&<q_{\rm wall}<1.011 .
\end{aligned}
\end{equation}
The safety factor therefore crosses unity on a regular torus, computed at $\psi\approx0.172$, on which every field line closes after one toroidal and one poloidal turn. There is no island chain and no current sheet. The wall lies inside the ball of the minimum-energy theorem, and $\lambda=2.25$ is below $x_1/r_b=2.528$, so E1 is the unique energy minimum in its wall. The energy rises by more than 8.8\% of the energy of any admissible change, so no ideal displacement inside the wall lowers it, the $m=n=1$ internal kink included. On a fine grid of points inside the wall, the cylindrical toroidal-field fraction $|B_\phi|/|\B|$ is at least $0.839$.

E1 is invariant under a half turn, so perturbations that keep this covariance have even toroidal mode numbers, and their lowest resonance at $q=1$ is $m=n=2$. The $q=1$ torus is intact because it belongs to the exact family of flux surfaces given by the first integral $\psi$, and the fixed-wall energy bound also applies to perturbations that break the half-turn symmetry.

\paragraph*{How low can $q(0)$ go?}
In a constant-$\lambda$ field $J_\parallel/B$ is uniform. In an axisymmetric near-axis or straight-cylinder estimate, an elliptic cross-section of elongation $\kappa$ with self-similar surfaces gives $q(0)=(2/\lambda R)(1+\kappa^2)/(2\kappa)$, and the circular Bessel solution gives $q(r)/q(0)=1-(\lambda r)^2/8+O((\lambda r)^4)$, but toroidicity and shaping can change the magnitude and the sign of the shear. We count an upward endpoint bracket when $q(0)<1<q_{\rm wall}$.

To see how far below unity $q(0)$ can fall, we scanned 500 geometries on a grid ($m_g\in\{0.2,0.35,0.5,0.65,0.8\}$, $R_0/K\in\{0.3,0.45,0.6,0.75,0.9\}$, $S_0/K\in\{0,0.25,0.5,0.75\}$ and $a/b\in\{1/4,1/2,1,2,4\}$ with $ab=1$), taking $\lambda$ up to the window and the wall up to the limit of the nested region. For walls with aspect ratio between 2.3 and 5 and nearly uniform sections around the torus, the lowest $q(0)$ with an upward endpoint bracket is about $0.94$, close to the DIII-D values quoted above. The lower-$q$ examples found in this scan have fat-torus geometry. The lowest, $q(0)=0.843$ ($m_g=0.8$, $R_0=0.45K$, $S_0=0$, $a=b=1$, $\lambda=2.819$ with window $2.822$, $\psi_w=2.468$), has $q$ rising to $q_{\rm wall}=1.15$ and crossing unity at $\psi\approx1.80$; its $q=1$ section has aspect ratio $1.5$ in the plane $y=0$ and $2.4$ in the plane $x=0$. This value is a minimum of the grid, not of the family: an independent random sample over the same ranges contains a nested configuration with $q(0)=0.838$.

\paragraph*{Shaping at fixed helicity.}
In Taylor-like relaxation, helicity and flux are conserved and the boundary is given; $\lambda$ follows. To ask whether shaping matters we therefore compare states at equal helicity-to-flux ratio. For a field with nested surfaces, with toroidal flux $\Phit$ and poloidal flux $\Phip$ measured from the axis, we use
\begin{equation}
k_T\equiv\frac{K}{\Phi_w^2}=2\int_0^1 x\,\iota\,dx,\qquad x=\Phit/\Phi_w ,
\label{eq:kT}
\end{equation}
where $K=\int\mathbf{A}\cdot\B\,dV-\Phi_w\oint_L\mathbf{A}\cdot d\mathbf{l}$, with $L$ a wall longitude of zero linking and the fluxes oriented along $\B$, is the helicity invariant of the bounded torus~\cite{Reiman1981,Taylor1986}, and $\iota=d\Phi_p/d\Phi_t$. The wall-circulation subtraction makes $K$ independent of the choice of interior vector potential; the associated wall term must also be retained in comparing magnetic energies~\cite{Reiman1980,Reiman1981,Taylor1986}. With the circulation of $\mathbf{A}$ along the magnetic axis set to zero, the loop term contributes $\Phi_w\Phi_{p,w}$. For a shearless profile $k_T=\iota$. With the axis circulation set to zero, the uncorrected volume integral divided by $\Phi_w^2$ is $k_T-\Phi_{p,w}/\Phi_w$. It vanishes for every shearless profile, whatever its $q$, so a near-zero value does not imply $q$ close to unity: among the 7500 configurations evaluated below, those with it within about $0.0008$ of zero have $q(0)$ from $0.81$ to $8.3$. Omitting the loop term therefore changes the conserved constraint used in the relaxation problem. We evaluated 7500 configurations (the 300 grid geometries with $a/b\le1$, five values of $\lambda$ up to the window and five walls up to the nested limit).

Figure~\ref{fig:shape} shows three configurations with $k_T\approx1.000$. Configuration (i), a nearly toroidally uniform section with $\kappa\approx1.3$, has $q$ rising from $0.952$ to $1.035$. Configuration (ii), whose section changes strongly around the torus (elongation about $1.7$ at $y=0$ and $2.9$ at $x=0$, triangularity about $0.03$ and $0.63$), rises from $0.983$ to $1.006$. Configuration (iii), an oblate, strongly three-dimensional section, has $q$ falling from $1.015$ to $0.996$ (negative shear). All three satisfy the window condition of (iv).

\begin{figure}[t]
\includegraphics[width=\columnwidth]{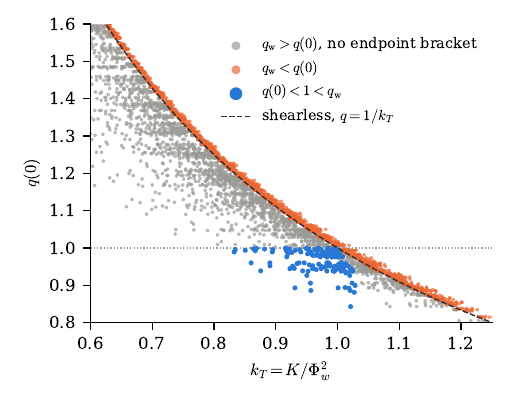}
\caption{Central safety factor against $k_T$ for the 7500 configurations of the sample, each satisfying the window condition of (iv); shown are $0.6<k_T<1.25$ and $0.8<q(0)<1.6$. Blue: $q(0)<1<q_{\rm wall}$. Orange: $q_{\rm wall}<q(0)$. Gray: $q_{\rm wall}>q(0)$, no endpoint bracket. Dashed: a shearless profile, $q=1/k_T$.}
\label{fig:q0k}
\end{figure}

Figure~\ref{fig:q0k} shows the whole set, with 152 upward and 27 downward endpoint brackets of $q=1$. For a profile with $q\geq q(0)$ throughout, $k_T$ is a weighted mean of $\iota$, so $q(0)\le1/k_T$ and the profile lies below the shearless line; a profile with $q\le q(0)$ throughout lies above it. All 152 upward brackets lie below the line. In this sample and in an independent random sample of 5000 configurations over the same ranges, the upward brackets cluster at $k_T$ of about $0.81$--$1.06$. At nearly equal $k_T$, wall shaping affects $q(0)$. The configurations of Fig.~\ref{fig:shape} have either sign of shear and, among the upward brackets, adding aspect ratio, elongation and triangularity to a linear fit in $k_T$ raises the explained variance of $q(0)$ from 9\% to 44\%. These numbers describe our two samples; other samplings of the family may give other values.

\paragraph*{A near-axisymmetric example.}
We also fitted the parameters towards a target shape, a DIII-D-like boundary (aspect ratio 2.7, elongation 1.8, triangularity 0.5), with $q(0)<1<q_{\rm wall}$ as a constraint. The fit gives $m_g=0.948$, $R_0=0.160K$, $S_0=0$, $a/b=1.011$ ($ab=1$), $\lambda=1.257$ and $\psi_w=2.407$. The fit is partial: in the planes $x=0$ and $y=0$ the wall has aspect ratio about 2.6 and triangularity 0.40--0.47, but elongation about 2.2 instead of 1.8. Using a geometric poloidal angle about the magnetic axis, the non-axisymmetric toroidal Fourier components of the wall coordinates $(R,Z)$ have a combined rms amplitude of about 1\% of the horizontal minor radius. The axis is planar and its radius varies by 0.6\%. Here $q$ rises from $0.99$ to $1.10$, and $\lambda$ is below the window. Minimum-energy states with $q(0)<1$ thus exist both far from and close to axisymmetry.

\paragraph*{Discussion.}
A fixed-wall Taylor state need not have $q(0)\geq1$. In a suitably shaped wall it can have $q(0)<1$ with nested surfaces through $q=1$, no island, and energy strictly below every other field with the same flux and helicity. If such a state models a relaxed core, its post-crash $q(0)$ depends on the conserved toroidal flux and helicity and the shape of the relaxing region, whereas Kadomtsev's model gives $q(0)\geq1$ and the model of Ref.~\onlinecite{Jardin2020} keeps it above unity. The KSTAR mode analysis favors $q(0)\geq1$ after the crash, while the DIII-D polarimetry finds $q(0)<1$ throughout the cycle. A test of the picture is to compare such states with measured $q$ profiles, in particular with the shear inside $q=1$ and with the persistence of the $q=1$ surface.

Several limits apply. The fields carry no pressure gradient, so the pressure-driven $1/1$ mode and the helical cores of Refs.~\onlinecite{Cooper2010,Brunetti2014} are outside the model. We do not give a dynamical path to these equilibria.

The main limit is partial relaxation: in a sawtooth only the region inside the mixing radius relaxes. A partial-relaxation model can treat this core as a relaxed solid torus at given toroidal flux and helicity; annular regions carry, in addition, an independent poloidal-flux constraint~\cite{Hudson2012,Qu2025}. Embedding an exact core in a global equilibrium requires an exterior field that satisfies the interface flux and pressure-balance conditions~\cite{Hole2006,Hudson2012}; $\lambda$ may jump across the interface but need not. The exact solutions given here are candidate relaxed cores and fixed-boundary benchmarks for that calculation.

Other exact force-free equilibria with nested toroidal magnetic surfaces will be reported elsewhere. They include a constant-$\lambda$ family on the confocal paraboloidal web with interior minima of $\iota$ and configurations containing two intact $\iota=2/3$ tori of opposite magnetic shear, and a convergent-series construction of variable-current force-free fields with closed lines and $\iota=1$ throughout a nested core.

\begin{acknowledgments}
This research was supported by the Department of Energy (Award No.~DE-SC0024548). The authors used AI coding assistants (OpenAI Codex and Anthropic Claude Code) for symbolic and numerical calculations, interval certificates, field-line tracing, parameter scans, literature searches and drafting. The authors set the research questions, checked the results and take responsibility for the scientific claims.
\end{acknowledgments}

\section*{Author declarations}
\subsection*{Conflict of interest}
The authors have no conflicts to disclose.
\section*{Data availability}
The data that support the findings of this study are available from the corresponding author upon reasonable request.


\begin{thebibliography}{99}
\bibitem{Grad1967} H. Grad, Phys. Fluids \textbf{10}, 137 (1967).
\bibitem{Grad1985} H. Grad, Int. J. Fusion Energy \textbf{3}, 33 (1985).
\bibitem{GLT2026} J. G\'omez-Serrano, L. Liehr, and M.~A. Taylor, arXiv:2609.24739 (2026).
\bibitem{Landreman2026} M. Landreman, arXiv:2609.26742 (2026).
\bibitem{Issan2026a} O. Issan, J. Citrin, M. Landreman, and B. Tracey, arXiv:2610.07304 (2026).
\bibitem{Issan2026b} O. Issan, L. Jos\'e, M. Landreman, J. Citrin, and B. Tracey, arXiv:2611.07672 (2026).
\bibitem{DeLacySato2026} K. de Lacy and N. Sato, arXiv:2611.16688 (2026).
\bibitem{EPPT2026} A. Enciso, J. Pe\~nafiel-Tom\'as, and D. Peralta-Salas, arXiv:2611.21574 (2026).
\bibitem{Sengupta2026vac} W. Sengupta, A. Bhattacharjee, T.~J. Klotz, A.~O. Brown, J.~W. Burby, V.~N. Duarte, S. Patil, and S. Buller, arXiv:2610.28466 (2026).
\bibitem{Lopez2026} O.~E. L\'opez, arXiv:2610.29921 (2026).
\bibitem{EncisoPeralta2015} A. Enciso and D. Peralta-Salas, Acta Math. \textbf{214}, 61 (2015).
\bibitem{EncisoLuquePeralta2020} A. Enciso, A. Luque, and D. Peralta-Salas, Adv. Math. \textbf{373}, 107328 (2020).
\bibitem{EncisoPeraltaRomaniega2023} A. Enciso, D. Peralta-Salas, and \'A. Romaniega, Forum Math. Sigma \textbf{11}, e56 (2023).
\bibitem{EncisoPeralta2016} A. Enciso and D. Peralta-Salas, Arch. Ration. Mech. Anal. \textbf{220}, 243 (2016).
\bibitem{ClellandKlotz2020} J.~N. Clelland and T. Klotz, Arch. Ration. Mech. Anal. \textbf{236}, 767 (2020).
\bibitem{Abe2022} K. Abe, J. Math. Phys. \textbf{63}, 041507 (2022).
\bibitem{PeraltaVaquero2024} D. Peralta-Salas and M. Vaquero, J. Math. Anal. Appl. \textbf{540}, 128632 (2024).
\bibitem{Taylor1974} J.~B. Taylor, Phys. Rev. Lett. \textbf{33}, 1139 (1974).
\bibitem{Taylor1986} J.~B. Taylor, Rev. Mod. Phys. \textbf{58}, 741 (1986).
\bibitem{Kadomtsev1975} B.~B. Kadomtsev, Sov. J. Plasma Phys. \textbf{1}, 389 (1975).
\bibitem{Soltwisch1988} H. Soltwisch, Rev. Sci. Instrum. \textbf{59}, 1599 (1988).
\bibitem{Yamada1994} M. Yamada, F.~M. Levinton, N. Pomphrey, R. Budny, J. Manickam, and Y. Nagayama, Phys. Plasmas \textbf{1}, 3269 (1994).
\bibitem{Wroblewski1993} D. Wroblewski and R.~T. Snider, Phys. Rev. Lett. \textbf{71}, 859 (1993).
\bibitem{Nam2018} Y.~B. Nam, J.~S. Ko, G.~H. Choe, Y. Bae, M.~J. Choi, W. Lee, G.~S. Yun, S. Jardin, and H.~K. Park, Nucl. Fusion \textbf{58}, 066009 (2018).
\bibitem{Benedett2024} T.~E. Benedett, J. Chen, D.~L. Brower, and W.~X. Ding, Plasma Phys. Control. Fusion \textbf{66}, 085009 (2024).
\bibitem{Park2016} H. Park, G. Yun, G. Choe, J. Ko, S. Jardin, W. Lee, Y. Nam, and Y. Bae, ``Validation of $q(0)\ge1.0$ in the MHD quiescent time after crash of the sawtooth instability in KSTAR,'' 26th IAEA Fusion Energy Conference, Kyoto (2016), EX/P4-3.
\bibitem{Jardin2020} S.~C. Jardin, I. Krebs, and N. Ferraro, Phys. Plasmas \textbf{27}, 032509 (2020).
\bibitem{Bhattacharjee1980} A. Bhattacharjee, R.~L. Dewar, and D.~A. Monticello, Phys. Rev. Lett. \textbf{45}, 347 (1980); \textbf{45}, 1217(E) (1980).
\bibitem{Taylor1975} J.~B. Taylor, in \textit{Proceedings of the 5th International Conference on Plasma Physics and Controlled Nuclear Fusion Research, Tokyo, Japan, 1974} (IAEA, Vienna, 1975), Vol.~I, p.~161.
\bibitem{Bhattacharjee1983} A. Bhattacharjee and J.~C. Wiley, Phys. Fluids \textbf{26}, 520 (1983).
\bibitem{Gimblett1994} C.~G. Gimblett and R.~J. Hastie, Plasma Phys. Control. Fusion \textbf{36}, 1439 (1994).
\bibitem{Qu2025} Z.~S. Qu, Y. Zhou, A. Kumar, J. Doak, J. Loizu, and M.~J. Hole, Phys. Plasmas \textbf{32}, 052503 (2025).
\bibitem{Reiman1980} A. Reiman, Phys. Fluids \textbf{23}, 230 (1980).
\bibitem{Reiman1981} A. Reiman, Phys. Fluids \textbf{24}, 956 (1981).
\bibitem{Whittaker1903} E.~T. Whittaker, Math. Ann. \textbf{57}, 333 (1903).
\bibitem{Bergman1929} S. Bergmann, Math. Ann. \textbf{101}, 534 (1929).
\bibitem{Gilbert1960} R.~P. Gilbert, Pacific J. Math. \textbf{10}, 1243 (1960).
\bibitem{Sengupta2026long} W. Sengupta, A. Bhattacharjee, T.~J. Klotz, and A.~O. Brown, ``Exact vacuum and force-free magnetic fields with nested flux surfaces: construction, singular limits and current sheets,'' in preparation (2026).
\bibitem{Cantarella2000} J. Cantarella, D. DeTurck, H. Gluck, and M. Teytel, Phys. Plasmas \textbf{7}, 2766 (2000).
\bibitem{Cooper2010} W.~A. Cooper, J.~P. Graves, A. Pochelon, O. Sauter, and L. Villard, Phys. Rev. Lett. \textbf{105}, 035003 (2010).
\bibitem{Brunetti2014} D. Brunetti, J.~P. Graves, W.~A. Cooper, and D. Terranova, Nucl. Fusion \textbf{54}, 064017 (2014).
\bibitem{Hole2006} M.~J. Hole, S.~R. Hudson, and R.~L. Dewar, J. Plasma Phys. \textbf{72}, 1167 (2006).
\bibitem{Hudson2012} S.~R. Hudson, R.~L. Dewar, G. Dennis, M.~J. Hole, M. McGann, G. von Nessi, and S. Lazerson, Phys. Plasmas \textbf{19}, 112502 (2012).
\end{thebibliography}
\end{document}